# Electron configuration of the substitutional nitrogen defect in diamond

**Jozsef Garai**

University of Debrecen, Hungary

It is suggested that the substitutional nitrogen in diamonds bonded to three of the surrounding carbon atoms instead of four. This proposed electron configuration of the defect is deduced from previous experiments and theoretical considerations. Notably, the 1344 cm$^{-1}$ band, characteristics of the substitutional Nitrogen, is independent of the isotopic change of Nitrogen but depend on the isotopic change of Carbon. The well established NV centre should not be stable if Nitrogen is bounded to four of the surrounding Carbon. Additional support comes from the substantially bigger size of the single substitutional nitrogen atom indicating loan pair electron. The proposed configuration of the substitutional Nitrogen was also tested by using a simple force constant model. Replacing force constant of C-N with 2/3 C-C:1/3 C=C reproduces the 1344 cm$^{-1}$ band.

**1. Introduction**  Nitrogen is the most important impurity in diamonds which changes the physical characteristics, like color and semi-conductivity. The importance of nitrogen is acknowledged by establishing a diamond classification based on the content and the type of the nitrogen defects [1]. There are more than 50 forms of nitrogenous defects that occur in diamonds. When a diamond forms, nitrogen atoms replace the carbon atoms, forming an isolated single impurity. All synthetic diamonds contain dominantly single nitrogen impurities. Thus knowing the physical properties of this defect is extremely important for not just science but industry as well.

There is a consensus that the substitutional nitrogen atom in the diamond structure forms four hybrid bonds with the neighboring carbon atoms [2-8]. Despite the simplicity of the substitutional N defect in a simple solid computational models seem to have problems to meet with experimental results [7, 9]. The widely accepted description of the substitutional nitrogen defects also contradicts with several experiments which are discussed here.

**2. The characteristic IR bands**  The characteristic IR bands of the substitutional nitrogen impurities are at 1130 cm$^{-1}$ and 1344 cm$^{-1}$. The strengths of these nitrogen-induced bands show excellent correlation. However, replacing the nitrogen-14 with nitrogen-15 results in a red shift of the 1130 cm$^{-1}$ band but no shift occurs for the 1344 cm$^{-1}$ band [8]. The unchanged frequency of the 1344 cm$^{-1}$ band indicates that the nitrogen atom do not participate in the vibration. The carbon atoms induced vibration of his band is further supported by the observed shift of the 1344 cm$^{-1}$ to 1292 cm$^{-1}$ when the host isotope is changed from carbon-12 to carbon-13. Assuming simple harmonic motion this shift is consistent with the frequency ratios of $(12/13)^{1/2}$ or 0.9607. If the nitrogen would be involved in the vibration then the ratio of the square root of the reduced mass is 0.979 for vibrations involving $^{14}$N-$^{13}$C and $^{14}$N-$^{12}$C bonds [10]. These isotope doping experiments indicates that the band observed at 1344 cm$^{-1}$ is induced by the vibration of the carbon atoms surrounding the substitutional N defect. The independent carbon induced vibration of the substitutional N defect can be explained by a non bonded carbon atom surrounding the substitutional nitrogen defect.

The bond energy correlates to the stretching force constant if the molecules are closely related with similar bonds [11]. Using linear correlation the stretching force constant for the C-C vibration [$k_{C-C}$] is calculated from the stretching force constant of the C-N vibration [$k_{C-N}$] as:

$$k_{C-C} = \frac{2 \times D(C-C) + D(C=C)}{3 \times D(N-C)} \times k_{C-N} \quad (1)$$

where D(C-C), D(C=C) and D(N-C) are the average bond energies. Using the values [12, 13] of D(C-C) = 347-348 KJmol$^{-1}$, D(C=C) = 611-612 KJmol$^{-1}$ and D(N-C) = 305 KJmol$^{-1}$ gives 1.426-1.429 for the constant multiplier.

It is assumed that the double bond of carbon resonates among the bonded three carbon atoms (2/3 C-C:1/3 C=C). Using a simple harmonic model gives a vibrational frequency induced by the carbon atoms of 1349/1351 cm$^{-1}$. This calculated value is in excellent agreement with the observed 1344 cm$^{-1}$. Thus the vibration of the 1344 cm$^{-1}$ band should be induced by the stretching vibration of the 2/3 C-C:1/3 C=C carbon atoms.

**3. Theoretical Considerations** Based on the spatial distribution of the sp$^3$ hybridized electrons Carbon atoms in the diamond structure have tetrahedron coordination. In order to fit into the diamond lattice the spatial distribution of the electrons of Nitrogen has to be the same. The electronic structure of Nitrogen is 2s$^2$ 2p$^3$. The three bonding domains of the p electrons and the one nonbonding domain of the loan electron pair electrons forms a tetrahedron skeleton structure which fits into the diamond lattice. Thus, the single nitrogen impurity in the diamond lattice should be bonded to three of the neighboring carbon atom but not to the fourth one [Fig. 1].

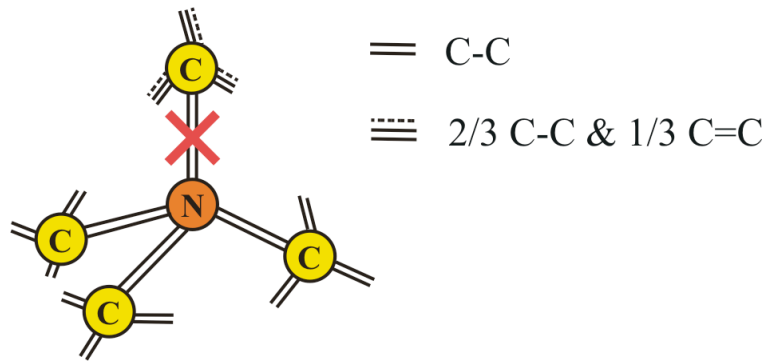

**Fig. 1** Schematic figure of the proposed bonding is shown.

It has been shown that the single substitutional nitrogen centres trap vacancies about eight times more efficiently than the substitutional nitrogen pairs [14]. If the nitrogen would be bonded to four carbon atoms then NV centre would not be stable. Thus the existence of these centers prove that the substitutional nitrogen is bonded to three of the surrounding carbon atoms instead of four.

It has been shown that the size of a single substitutional nitrogen atom in diamond is 1.41 times that of the carbon atom it replaces [15]. This substantially bigger size of the nitrogen is also consistent with the existence of an additional lone pair electron.

**4. Conclusions** Based on the stability of the NV centre, on the independence of the 1344 cm$^{-1}$ band from the nitrogen isotopes and on the dependence of the carbon isotope, on the substantially bigger size of the single substitutional nitrogen and on reproduction of the 1344 cm$^{-1}$ band using 2/3 C-C:1/3 C=C force constant instead of C-N, it is suggested that the single nitrogen impurity in diamonds is bonded to three of the surrounding carbon atoms.